\documentclass[aps,pra,reprint,groupedaddress,footinbib,longbibliography]{revtex4-2}

\usepackage{graphicx}
\usepackage{dcolumn}
\usepackage{bm}
\usepackage{amsmath,amssymb}
\usepackage{bbold}
\usepackage[cal=boondoxo]{mathalfa}
\allowdisplaybreaks
\usepackage[bookmarksnumbered,bookmarksopen]{hyperref}
\hypersetup{
   colorlinks=true,       
   linkcolor=red,         
   citecolor=blue,        
}
\usepackage{url}
\usepackage{color}

\newcommand* {\vek}[1]{{\bm{\mathrm{#1}}}}
\newcommand* {\kk}{\vek{k}}
\newcommand* {\kkop}{\hat{\vek{k}}}

\newcommand* {\rr}{\vek{r}}
\newcommand* {\ee}{\mathrm{e}}

\let\myRe\Re
\let\myIm\Im
\renewcommand{\Re}{\myRe\mathrm{e}\,}
\renewcommand{\Im}{\myIm\mathrm{m}\,}

\graphicspath{{.}{./FIG/}}

\begin{document}


\title{Coexistence of topological and nontopological
Fermi-superfluid phases}

\author{K. Thompson}
\affiliation{Dodd-Walls Centre for Photonic and Quantum
Technologies, School of Chemical and Physical Sciences,
Victoria University of Wellington, PO Box 600, Wellington
6140, New Zealand}

\author{U. Z\"ulicke}
\email{uli.zuelicke@vuw.ac.nz}
\affiliation{Dodd-Walls Centre for Photonic and Quantum
Technologies, School of Chemical and Physical Sciences,
Victoria University of Wellington, PO Box 600, Wellington
6140, New Zealand}

\author{J. Brand}
\affiliation{Dodd-Walls Centre for Photonic and Quantum
Technologies, Centre for Theoretical Chemistry and Physics,
New Zealand Institute for Advanced Study, Massey University,
Private Bag 902104, North Shore, Auckland 0745, New Zealand}

\date{\today}

\begin{abstract}
The two-dimensional spin-imbalanced Fermi gas subject to
\textit{s}-wave pairing and spin-orbit coupling is considered a
promising platform for realizing a topological
chiral-\textit{p}-wave superfluid. In the BCS limit of
\textit{s}-wave pairing, i.e., when Cooper pairs are only
weakly bound, the system enters the topological phase via a
second-order transition driven by increasing the Zeeman
spin-splitting energy. Stronger attractive two-particle
interactions cause the system to undergo the BCS--BEC
crossover, in the course of which the topological transition
becomes first-order. As a result, topological and
nontopological superfluids coexist in spatially separated
domains in an extended region of phase space spanned by the
strength of \textit{s}-wave interactions and the Zeeman energy.
Here we investigate this phase-coexistence region theoretically
using a zero-temperature mean-field approach. Exact numerical
results are presented to illustrate basic physical
characteristics of the coexisting phases and to validate an
approximate analytical description derived for weak spin-orbit
coupling. Besides extending our current understanding of
spin-imbalanced superfluid Fermi systems, the present approach
also provides a platform for future studies of unconventional
Majorana excitations that, according to topology, should be
present at the internal interface between coexisting
topological and nontopological superfluid parts of the system.
\end{abstract}

\maketitle

\section{\label{sec:intro}Introduction}

The superfluidity of polarized, i.e., species-imbalanced Fermi
gases underpins a wide range of topics focused on in current
research~\cite{Chevy2010,Radzihovsky2010}. Pioneering works
were motivated by the desire to understand how Zeeman spin
splitting affects \textit{s}-wave pairing of electrons in
metals~\cite{Chandrasekhar1962,Clogston1962,Sarma1963}. Since
then, the possibility to tune the attractive two-particle
interaction strength in ultracold atom gases using Feshbach
resonances~\cite{Chin2010} has opened up the opportunity to
explore the BCS--BEC crossover~\cite{Leggett1980a,Nozieres1985,
Chen2005,Strinati2018} in polarized Fermi gases, revealing
interesting features of their phase diagram~\cite{Bedaque2003,
Carlson2005,Sheehy2006,Son2006,Parish2007,He2008,Du2009,
Olsen2015}. The recent advent of tunable spin-orbit couplings
in both condensed-matter and ultracold-atom realizations of the
spin-$1/2$ Fermi gas~\cite{Galitski2013,Manchon2015} has
galvanized interest in studying the combined effects of spin
polarization and spin-orbit coupling on \textit{s}-wave
pairing~\cite{Zhang2008,Sau2010,Alicea2010,Sato2010,
Kubasiak2010,Zhu2011,Iskin2011,Gong2011,Tewari2011,Yi2011,
Zhou2011,Yang2012,Jiang2011,Liu2012,Yang2012a,Liu2012a,Seo2012,
Liu2012b,He2012,He2013,Qu2013,Zhang2013,Xu2014,Liu2015,Zou2016,
Brand2018,Thompson2020}, especially the emergence of
topological superfluidity~\cite{Loder2015,Beenakker2016,
Sato2017}.

\begin{figure}[t]
\includegraphics[width=\columnwidth]{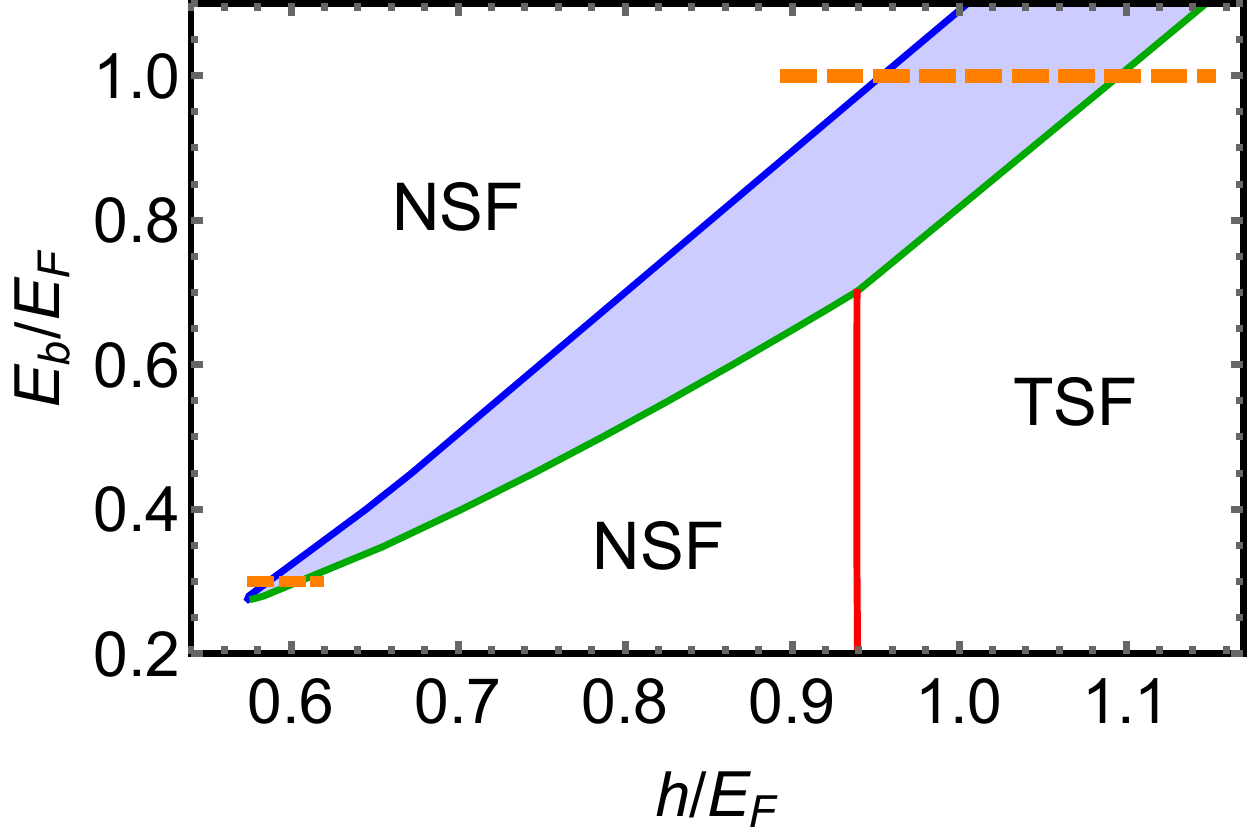}
\caption{\label{fig:Diag}%
Mean-field phase diagram for a spin-orbit-coupled 2D Fermi gas
at zero temperature and fixed density $n = m E_\mathrm{F}/(\pi
\hbar^2) \equiv k_\mathrm{F}^2/(2\pi)$ in the parameter space
of attractive-interaction strength (quantified in terms of the
two-particle binding energy $E_\mathrm{b}$ in vacuum without
spin-orbit coupling and Zeeman spin splitting) and Zeeman
energy $h$. The blue and green curves correspond to the
critical Zeeman energies $h_<$ and $h_>$ that delimit the
first-order phase-transition region $h_< < h < h_>$, indicated
by light-blue shading. In this region, homogeneous ground-state
phases are not possible and coexistence between domains of
different superfluid phases is expected. The second-order
transition between nontopological-superfluid (NSF) and
topological-superfluid (TSF) phases occurs at the red curve
where $h=h_\mathrm{c}$. Orange dashed lines illustrate the
parameter range for which further data are shown in
Fig.~\ref{fig:Coex}. Results presented in the figure have been
calculated for the spin-orbit-coupling strength $\lambda =
0.35\, E_\mathrm{F}/k_\mathrm{F}$.}
\end{figure}

Our present study is focused on zero-temperature properties of
the two-dimensional (2D) spin-$1/2$ Fermi gas with attractive
interactions. The interaction strength can be parameterized in
terms of the energy $E_\mathrm{b}$ of the two-particle bound
state in vacuum without spin-orbit coupling and Zeeman
splitting, which exists in a 2D system at any nonzero strength
of \textit{s}-wave interactions \cite{Randeria1990,
Levinsen2015}. In the absence of spin-orbit coupling, raising
the Zeeman energy splitting $2 h$ between opposite-spin states
drives a first-order transition from the \textit{s}-wave
superfluid phase to the normal phase, regardless of the
magnitude of $E_\mathrm{b}$~\cite{He2008}. Introducing
spin-orbit coupling drastically changes the $E_\mathrm{b}$-$h$
phase diagram~\cite{Yi2011,Zhou2011,Yang2012,He2013,
Thompson2020}: the first-order transition is now between two
superfluid phases and occurs only above a critical value
$E_\mathrm{b}^\mathrm{(c)}$. In addition, homogeneous
superfluid phases realize a topological superfluid (TSF) when
$h$ is larger than a critical value
$h_\mathrm{c}$~\cite{Sau2010,Alicea2010,Sato2010} determined by
the chemical potential $\mu$ and the \textit{s}-wave
pair-potential magnitude $|\Delta|$ via
\begin{equation}\label{eq:topCrit}
h_\mathrm{c} = \sqrt{\mu^2 + |\Delta|^2} \quad .
\end{equation}
These features are illustrated in Fig.~\ref{fig:Diag} for a 2D
Fermi gas with fixed total particle density $n$, using the
scales $E_\mathrm{F}=\pi\hbar^2n/m$ and $k_\mathrm{F}=\sqrt{2
\pi n}$ as energy and wave-vector units,
respectively~\footnote{Figure~\ref{fig:Diag} shows results
obtained for a system with relatively weak spin-orbit coupling.
For comparison, see Fig.~2(a) from Ref.~\cite{He2008} depicting
the case without spin-orbit coupling and Fig.~1 from
Ref.~\cite{Thompson2020} for situations with moderately large
spin-orbit-coupling magnitude.}. 
The shaded region shown in the figure corresponds to parameter
combinations for which homogeneous ground-state phases cannot
exist. Instead, the ground state of the system in this region
will form domains of different coexisting homogeneous phases,
similar to the familiar liquid-gas equilibrium in the
thermodynamics of real gases \cite{Callen1985}. Interestingly,
the curve $h_\mathrm{c}(E_\mathrm{b})$ associated with the
topological transition generically crosses into the
phase-coexistence region at some binding energy $\ge
E_\mathrm{b}^\mathrm{(c)}$. As a result, the first-order
superfluid-superfluid transition becomes intertwined with the
topological transition and coexistence now occurs between a
nontopological superfluid (NSF) and a TSF~\cite{Yang2012,
Thompson2020}. A conceptually different scenario for TSF--NSF
coexistence was proposed for trapped Fermi
gases~\cite{Zhou2011,Yang2012a,Liu2012,Liu2012a,Xu2014} where
the trap-potential-induced spatial variation of $\mu$ and
$\Delta$ causes $h_\mathrm{c}$ to be position-dependent, thus
creating the possibility for TSF and NSF regions to exist
simultaneously within a trap at fixed Zeeman energy $h$.

Here we provide a detailed theoretical analysis of the
first-order phase-coexistence region that has previously been
mapped in mean-field phase diagrams~\cite{Yang2012,
Thompson2020}. In particular, we obtain the chemical potential
as well as the pair-potential magnitudes of coexisting
superfluid phases as functions of the Zeeman energy $h$, the
spin-orbit-coupling strength $\lambda$, and the interaction
strength parameterized by $E_\mathrm{b}$. Systematic physical
insight is gained from approximate analytical expressions that
are derived assuming small spin-orbit coupling ($\lambda
k_\mathrm{F} \ll h$) and validated by comparison with exact
numerical results.

Experimental realizations of a TSF are vigorously
pursued~\cite{Beenakker2016,Meng2016,Lutchyn2018,Frolov2020,
Flensberg2021} because such systems host unconventional
zero-energy excitations~\cite{Kopnin1991,Volovik1999,Read2000,
Ivanov2001,Fu2008} that mimic properties of Majorana
fermions~\cite{Elliott2015} and could be used as building
blocks for fault-tolerant quantum bits~\cite{DasSarma2015}.
Such Majorana excitations occur generically at a TSF's
boundaries with vacuum or other nontopological
matter~\cite{Jackiw1976,Sato2017} and can thus be expected to
emerge also at the interface between the TSF coexisting with a
NSF in the first-order transition region explored in our
present work. The detailed understanding of the system in the
coexistence regime developed in this Article can inform a
realistic theoretical description~\cite{Thompson202x} of
Andreev bound states~\cite{Sauls2018} localized at the TSF--NSF
interface. Treating such an unconventional version of a
superfluid-superfluid (SS$^\prime$) hybrid structure where
pair-potential magnitudes are different on opposite sides
requires generalization of models applied previously to study
Josephson junctions~\cite{Kulik1969,Furusaki1991,Beenakker1991,
Bagwell1992,Spuntarelli2010}, solitons~\cite{Antezza2007,
Liao2011,Efimkin2015} and vortices~\cite{Mizushima2018}.

We present a detailed study of the first-order
phase-coexistence region in 2D Fermi superfluids with Zeeman
spin splitting and spin-orbit coupling in
Sec.~\ref{sec:PhaseCoex}. Exact numerical and approximate
analytical results are presented and compared with each other,
followed by a detailed discussion of their physical meaning and
implications. Section~\ref{sec:concl} contains our conclusions
and an outlook on their application. Mathematical details of
the derivations yielding our approximate analytical results are
provided in the Appendix.

\section{\label{sec:PhaseCoex}First-order superfluid-superfluid
transition in a spin-imbalanced 2D Fermi gas with spin-orbit
coupling}

We utilize the spin-resolved version of Bogoliubov--de~Gennes
(BdG) theory~\cite{deGennes1989,Ketterson1999} applicable to
the description of unconventional~\cite{Sigrist1991} or
noncentrosymmetric~\cite{Eschrig2012,Smidman2017} superfluids.
It is based on solving the general time-independent BdG
equation
\begin{equation}\label{eq:genBdG}
\begin{pmatrix}
\hat{H}_0(\kkop) & \hat{\Delta}(\kkop) \\[5pt] -\big[
\hat{\Delta}(\kkop)\big]^\ast & - \big[ \hat{H}_0(\kkop)
\big]^\ast \end{pmatrix} \begin{pmatrix} u(\rr) \\[5pt] v(\rr)
\end{pmatrix} = E \begin{pmatrix} u(\rr) \\[5pt] v(\rr)
\end{pmatrix}
\end{equation}
to obtain Bogoliubov-quasiparticle energies $E$ and associated
Nambu eigenspinors $(u, v)^T \equiv (u_\uparrow,u_\downarrow,
v_\uparrow, v_\downarrow)^T$ in the representation of 2D
position vector $\rr \equiv (x,y)$. Here and in the following,
the superscripts '$\ast$', '$T$', and '$\dagger$' are used to
denote complex conjugation, transposition in spinor space, and
Hermitian conjugation, respectively. The single-particle
Hamiltonian $\hat H_0(\kkop)$ and pair potential $\hat{\Delta}
(\kkop)$ are $2\times 2$ matrices in spin-$1/2$ space, which
for our system of interest are given by
\begin{subequations}
\begin{eqnarray}
\hat{H}_0(\kkop) &=& \left( \epsilon_{\kkop} - \mu \right)
\sigma_0 - h\, \sigma_z + \lambda_{\kkop}\, \sigma_+ + \big[
\lambda_{\kkop} \big]^\dagger\, \sigma_- \,\, , \quad \\[5pt]
\hat{\Delta}(\kkop) &=& -i\, \Delta\, \sigma_y \,\, .
\end{eqnarray}
\end{subequations}
Here we utilize the standard notation where the $\sigma_j$ with
$j=x,y,z$ denote Pauli matrices in spin-1/2 space, and
$\sigma_0$ is the identity operator in that subspace.
Furthermore, the combinations $\sigma_\pm = (\sigma_x \pm i
\sigma_y)/2$ are ladder operators for the eigenstates of
$\sigma_z$, $\kkop \equiv (\hat{k}_x, \hat{k}_y) = (-i
\partial_x, -i \partial_y)$ is the operator for the 2D wave
vector, $h$ denotes the Zeeman energy, and $\Delta$ is the, in
general, complex-number-valued \textit{s}-wave pair potential.
To be specific, we assume a quadratic single-particle
dispersion $\epsilon_{\kkop} = \hbar^2(\hat{k}_x^2 +
\hat{k}_y^2)/(2m)$ and Rashba-type~\cite{Bychkov1984}
spin-orbit coupling $\lambda_{\kkop} = \lambda (i \hat{k}_x +
\hat{k}_y)$ in the following, but our results apply more
generally.

For a homogeneous system, the Nambu eigenspinors are plane
waves in 2D coordinate ($\rr$) space, i.e.,
\begin{equation}
\begin{pmatrix} u(\rr) \\[5pt] v(\rr) \end{pmatrix} =
\begin{pmatrix} u_\kk \\[5pt] v_\kk \end{pmatrix}
\ee^{i\kk\cdot\rr} \quad ,
\end{equation}
with $\kk$ indicating 2D-wave-vector eigenvalues. The
associated quasiparticle eigenenergies are grouped into
four bands with dispersions $E_{\kk\alpha,\eta}$ with $\alpha
\in \{+,-\}$ and $\eta \in \{<,>\}$, given explicitly
by~\cite{Yi2011,Zhou2011}
\begin{widetext}
\begin{equation}\label{eq:qpSpec}
E_{\kk\alpha,<(>)} = \alpha \sqrt{(\epsilon_\kk -\mu)^2 +
|\Delta|^2 + h^2 + |\lambda_\kk|^2 \substack{-\\(+)} 2
\sqrt{(\epsilon_\kk -\mu)^2 ( h^2 + |\lambda_\kk|^2) +
|\Delta|^2 h^2}} \quad .
\end{equation}
\end{widetext}
In terms of these, the grand-canonical ground-state energy
density can be obtained via~\cite{Thompson2020}
\begin{eqnarray}\label{eq:gsEn}
&& \mathcal{E}_\mathrm{gs}^\mathrm{(MF)}(|\Delta|, \mu) =
\nonumber \\ && \hspace{1cm} \frac{1}{A}\sum_\kk \Big(
\frac{|\Delta|^2}{2 \epsilon_\kk + E_\mathrm{b}} +\epsilon_\kk
-\mu -\frac{1}{2} \sum_\eta E_{\kk+,\eta} \Big)\, , \quad
\end{eqnarray}
where $A$ denotes the area occupied by the 2D Fermi gas, and
$E_\mathrm{b}>0$ is the magnitude of the two-particle
bound-state energy in vacuum, for zero spin-orbit coupling and
also with no Zeeman spin splitting, introduced already in
Sec.~\ref{sec:intro}. A complete description of the homogeneous
Fermi superfluid at fixed density $n$ requires self-consistent
determination of the chemical potential $\mu$ and
pair-potential magnitude $|\Delta|$ such that the conditions
\begin{subequations}
\begin{eqnarray}\label{eq:minEgs}
\frac{\partial \mathcal{E}_\mathrm{gs}^\mathrm{(MF)}}{\partial
|\Delta|} &=& 0 \quad , \\[0.1cm]
\frac{\partial \mathcal{E}_\mathrm{gs}^\mathrm{(MF)}}{\partial
\mu} &=& - n \label{eq:denEgs}
\end{eqnarray}
\end{subequations}
are fulfilled, ensuring also that
$\mathcal{E}_\mathrm{gs}^\mathrm{(MF)}(|\Delta|, \mu)$ taken as
a function of $|\Delta|$ at fixed $\mu$ has a \emph{global\/}
minimum~\cite{Sheehy2007a}.

\begin{figure}[b]
\includegraphics[width=\columnwidth]{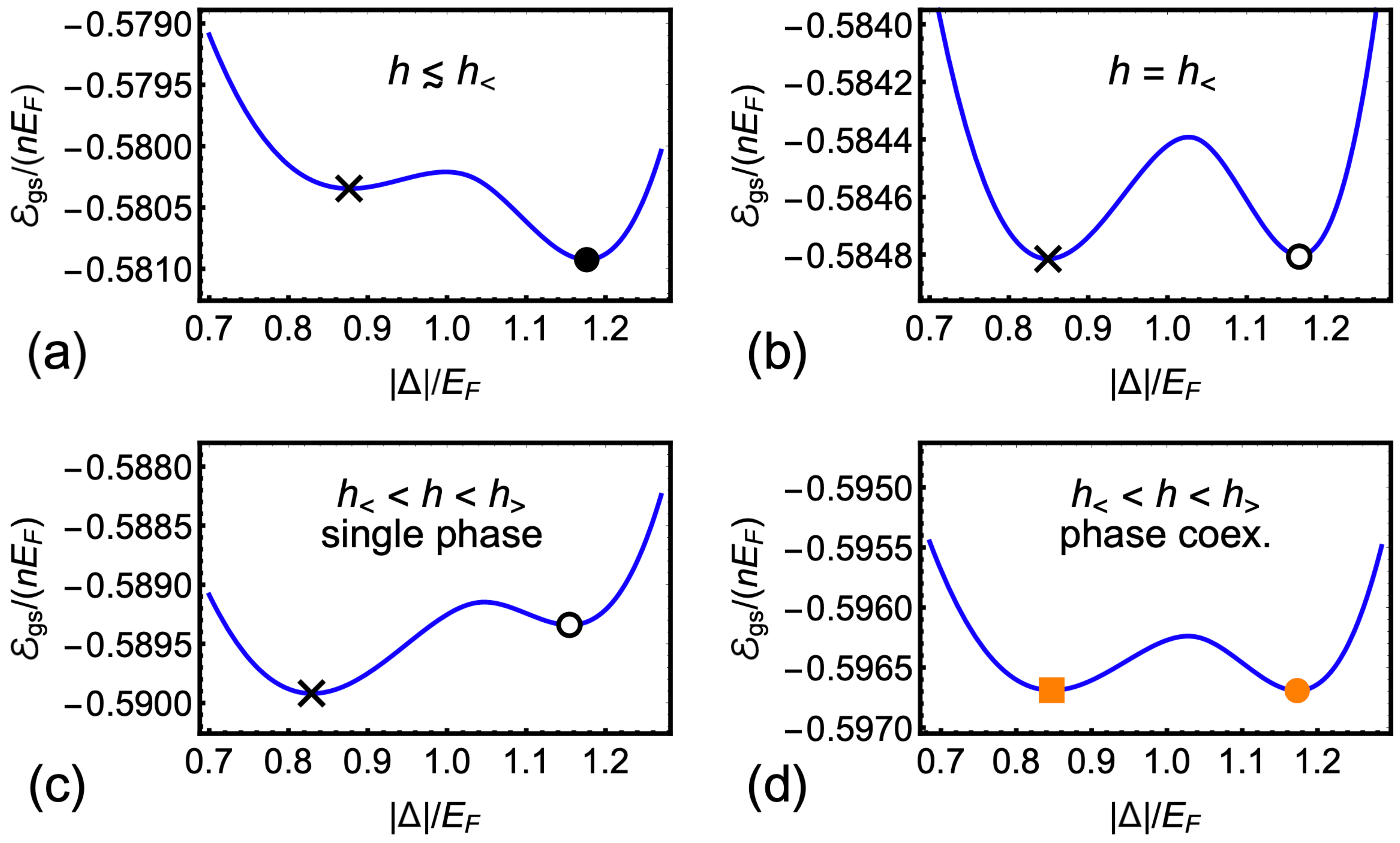}
\caption{\label{fig:EgsVsDelta}%
Evolution of minima in the mean-field ground-state energy
density $\mathcal{E}_\mathrm{gs}^\mathrm{(MF)}(|\Delta|, \mu)$,
taken as a function of the pair-potential magnitude $|\Delta|$
for fixed chemical potential $\mu$, when the Zeeman energy $h$
is increased beyond $h_<$. For $h\lesssim h_<$ [panel (a)], the
global minimum (marked by the filled black circle) is at the
value of $|\Delta|$ that satisfies the fixed-density condition
Eq.~(\ref{eq:denEgs}) with the given value of $\mu$, thus
representing a valid single-phase equilibrium ground state of
the system. As $h$ is increased, the energy difference between
the global minimum and another (local) minimum (marked with a
cross) reduces and, for $h=h_<$, both minima are degenerate
[panel (b)]. For $h_< < h < h_>$, the minimum associated with
the self-consistent value of $|\Delta|$ [indicated by the empty
black circle in panel (c)] then ceases to be the global
minimum. It therefore cannot any longer represent a viable
equilibrium ground state but remains a possible metastable
state. Adjusting $\mu$ to maintain degeneracy between the two
minima of $\mathcal{E}_\mathrm{gs}^\mathrm{(MF)}$ [panel (d);
see Eq.~(\ref{eq:coexCond})] determines the pair potentials
$|\Delta_\mathrm{s}|$ and $|\Delta_\mathrm{w}|$ (marked by the
filled orange circle and square, respectively) of coexisting
superfluid phases. Plotted data are for $E_\mathrm{b}=1.0\,
E_\mathrm{F}$, $\lambda=0.75\, E_\mathrm{F}/k_\mathrm{F}$ (all
panels) and $h=1.060\, E_\mathrm{F}$ [panel (a)], $1.064\,
E_\mathrm{F}$ [panel (b)], $1.068\, E_\mathrm{F}$ [panels (c)
and (d)].}
\end{figure}

As illustrated in the phase diagram shown in
Fig.~\ref{fig:Diag}, a homogeneous superfluid ground state is
found for the 2D Fermi gas with spin-orbit coupling except for
values of $h$ within a region $h_<(E_\mathrm{b}) < h <
h_>(E_\mathrm{b})$. This is because, for $h=h_<$ ($h_>$), the
global minimum of $\mathcal{E}_\mathrm{gs}^\mathrm{(MF)}
(|\Delta|, \mu)$ at the self-consistently determined value of
$|\Delta|$ becomes degenerate with a local minimum that is
present for $h\lesssim h_<$ ($h \gtrsim h_>$) and, when $h_< <
h < h_>$, the self-consistent $|\Delta|$ is no longer
associated with the global minimum of the ground-state energy
as required by thermodynamic stability but rather corresponds
to a local minimum or even a maximum. Panels (a)--(c) of
Fig.~\ref{fig:EgsVsDelta} illustrate this behavior of the
homogeneous, i.e., single-phase ground-state energy as the
Zeeman energy is raised beyond $h_<$. The impossibility to
realize a homogeneous ground state signals the breakdown of the
single-phase description for the system in this parameter
range. Instead, coexistence of two phases having the same value
of $\mu$ but different densities and pair-potential magnitudes
has to be assumed~\cite{Sheehy2007}, in the spirit of the
familiar treatment of the liquid-gas phase
transition~\cite{Callen1985}. Such a first-order phase
transition and associated coexistence region occur generically
in our system of interest when the two-particle \textit{s}-wave
binding energy $E_\mathrm{b}$ exceeds a critical value
$E_\mathrm{b}^\mathrm{(c)}$. As our previous systematic study
has shown~\cite{Thompson2020} [see Fig.~8(a) in that article],
$E_\mathrm{b}^\mathrm{(c)}$ increases monotonically with
increasing spin-orbit-coupling strength $\lambda$. This
explains the 'disappearance' of the first-order phase
transition noted in earlier works~\cite{Zhou2011,Tewari2011,
He2013} when $\lambda$ was increased while keeping
$E_\mathrm{b}$ fixed.

\begin{figure*}
\includegraphics[width=\textwidth]{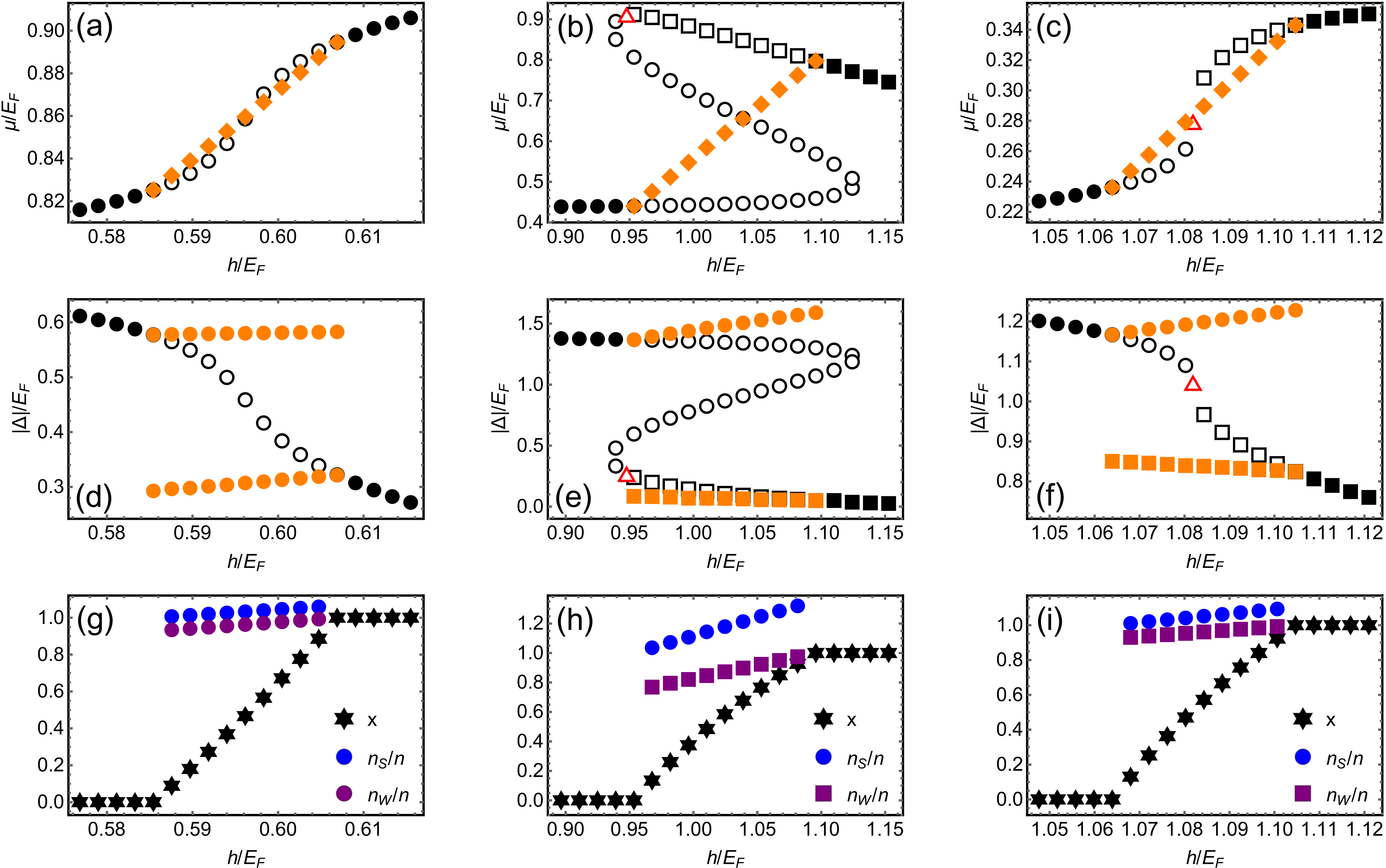}
\caption{\label{fig:Coex}%
First-order phase transition driven by the Zeeman
spin-splitting energy $h$ in a 2D Fermi gas, with fixed total
particle density $n$, subject to \textit{s}-wave pairing and
linear-in-momentum spin-orbit coupling. The system has a
single-phase superfluid ground state for $h < h_<$ and $h >
h_>$, with corresponding values for the chemical potential
$\mu$ [panels (a)--(c)] and the pair-potential magnitude
$|\Delta|$ [(panels (d)--(f)] indicated by filled black circles
(squares) if that state is a nontopological (a topological)
superfluid; i.e., $h<(>) \sqrt{\mu^2 + |\Delta|^2}$. The
transition region $h_< < h < h_>$ features coexistence of two
superfluid phases, one of which (the one with larger
$|\Delta|$) is always nontopological, whereas the other is
either nontopological or topological. The values of $\mu$ (of
$|\Delta|$) for the coexisting phases are indicated by filled
orange diamonds (filled orange circles or squares indicating
their nontopological or topological character, respectively).
Panels (g)--(i) show plots of the densities of coexisting
superfluid phases, together with the density fraction $x$ of
the phase with smaller $|\Delta|$, as a function of $h$.
Results shown in the left (middle, right) column of panels
have been calculated assuming a spin-orbit-coupling strength
$\lambda = 0.35\, E_\mathrm{F}/k_\mathrm{F}$ ($0.35\,
E_\mathrm{F}/k_\mathrm{F}$, $0.75\, E_\mathrm{F}/k_\mathrm{F}$)
and a two-particle binding energy $E_\mathrm{b} = 0.30\,
E_\mathrm{F}$ ($1.0\, E_\mathrm{F}$, $1.0\, E_\mathrm{F}$),
with $E_\mathrm{F}\equiv\hbar^2 k_\mathrm{F}^2/(2 m) = \pi
\hbar^2 n/m$. Empty circles (squares, triangles) indicate
values for $\mu$ and $|\Delta|$ that represent self-consistent
homogeneous nontopological (topological, critical) superfluid
states at the given $h$ but are not associated with a global
minimum of the ground-state energy, such as in the situation
depicted in Fig.~\ref{fig:EgsVsDelta}(c). These states are
metastable and, hence, will not be realized in equilibrium.}
\end{figure*}

We proceed to describe details of the theoretical approach and
present relevant results. In the phase-coexistence region, the
value of the chemical potential is determined via the
requirement~\cite{Sheehy2007} that the two minima, appearing at
pair-potential magnitudes $|\Delta_\mathrm{s}|$ and
$|\Delta_\mathrm{w}|$, respectively, in the ground-state energy
density taken as a function of $|\Delta|$ at fixed $\mu$ are
degenerate, i.e.,
\begin{equation}\label{eq:coexCond}
\mathcal{E}_\mathrm{gs}^\mathrm{(MF)}(|\Delta_\mathrm{s}|, \mu)
= \mathcal{E}_\mathrm{gs}^\mathrm{(MF)}(|\Delta_\mathrm{w}|,
\mu) \quad .
\end{equation}
The condition (\ref{eq:coexCond}) replaces the one formulated
in Eq.~(\ref{eq:denEgs}) for the homogeneous single-phase
ground state. Figure~\ref{fig:EgsVsDelta}(d) depicts the
situation that emerges from the one shown in panel~(c) after
$\mu$ is adjusted to ensure (\ref{eq:coexCond}) is satisfied.
To be specific, we assume $|\Delta_\mathrm{s}| >
|\Delta_\mathrm{w}|$, thus associating $|\Delta_\mathrm{s}|$
and $|\Delta_\mathrm{w}|$ with the strong and weak superfluids
in the coexistence state, respectively. Using these values as
input for the homogeneous-single-phase relation
(\ref{eq:denEgs}) yields the particle densities in the two
coexisting phases;
\begin{equation}
n_\mathrm{s,w} = -\left. \frac{\partial
\mathcal{E}_\mathrm{gs}^\mathrm{(MF)}}{\partial \mu}
\right|_{|\Delta|=|\Delta_\mathrm{s,w}|} \quad ,
\end{equation}
enabling us to determine the proportion $0\le x\le 1$ of the
weak superfluid in the coexistence state from the modified
fixed-density condition~\cite{Sheehy2007,Yi2011}
\begin{equation}
n = (1-x)\, n_\mathrm{s} + x\, n_\mathrm{w} \quad .
\end{equation}
Results obtained from this procedure for three representative
parameter combinations are shown in Fig.~\ref{fig:Coex}.

The strong superfluid phase turns out to be always
nontopological, i.e., $h<\sqrt{\mu^2 + |\Delta_\mathrm{s}|^2}$
for $h_< < h < h_>$. In contrast, the weak superfluid is
nontopological only for small-enough $E_\mathrm{b}$ and
not-too-large $\lambda$ --- specifically $\lambda\lesssim 0.7\,
E_\mathrm{F}/k_\mathrm{F}$~\cite{Thompson2020}. The left column
of panels in Fig.~\ref{fig:Coex} [Figs.~\ref{fig:Coex}(a,d,g)]
depicts such a situation. Conversely, when  $E_\mathrm{b}/
E_\mathrm{F}$ is above a critical value [as is the case for the
parameter combination used to calculate results presented in
the middle column, i.e., panels (b), (e) and (h) of
Fig.~\ref{fig:Coex}], or when $\lambda \ge 0.7\, E_\mathrm{F}/
k_\mathrm{F}$ [as for Fig.~\ref{fig:Coex}(c,f,i)], $h>\sqrt{
\mu^2+|\Delta_\mathrm{w}|^2}$ throughout the phase-coexistence
region and the weak-superfluid phase is topological.

Across the first order phase transition, we find the density
$n_\mathrm{w}$ of the weak superfluid phase to be always
smaller than the density $n_\mathrm{s}$ of the
strong-superfluid phase, regardless of whether the weak
superfluid phase is a TSF or an NSF. It is thus possible to
predict the spatial distribution of coexisting phases in
physically realistic situations when the system is subject to
finite potential gradients, e.g., because it is in a trap or
subject to gravity. According to the local density
approximation, an inhomogeneous external potential gives rise
to a spatially varying effective chemical potential of local
equilibrium, which is maximised at the minimum of the
potential. Since the chemical potential is a monotonously
growing function of the density (a stability condition of the
homogeneous gas), it follows that the high-density strong
superfluid phase (always NSF) will accumulate in the central
region of a trapping potential under coexistence conditions,
while the low-density weak superfluid phase (TSF or NSF) will
occupy the shallower region of the potential. This is
consistent with the results obtained from the local density
approximation in Ref.~\cite{Zhou2011}. The expected layering of
phases is indeed the configuration found in numerical
mean-field studies of spin-orbit-coupled Fermi superfluids in
2D~\cite{Liu2012} and 1D~\cite{Liu2012a,Xu2014} traps. In the
case that the weak superfluid is topological, the internal
boundary between the two coexisting phases constitutes a
TSF--NSF interface. More generally, a boundary between the two
coexisting phases can be expected to form under any
circumstances and, to minimize the associated energy
cost~\cite{DeSilva2006,Caldas2007,Haque2007,Baur2009}, will
typically have a simply connected shape and shortest-possible
length as allowed by sample geometry.

Empirical observation suggests that the $\mu(h)$ curve in the
coexistence region exhibits features similar to the familiar
Maxwell construction~\cite{Callen1985}. This observation
emerges from a comparison with the $\mu$ values of metastable
single-phase states [shown as empty symbols in
Figs.~\ref{fig:Coex}(a,b,c)] that correspond to only local, not
global minima in the $|\Delta|$-dependence of
$\mathcal{E}_\mathrm{gs}^\mathrm{(MF)}(|\Delta|, \mu)$ at fixed
$\mu$, such as the one indicated by the empty circle in
Fig.~\ref{fig:EgsVsDelta}(c). We explored metastable-state
characteristics in our earlier work~\cite{Thompson2020},
including the possibility to have more than a single one of
these at a given value of $h$ in certain parameter regimes, as
is the case for the situation depicted in
Figs.~\ref{fig:Coex}(b,e). Thus the system lends itself to
further experimental and theoretical study of processes akin to
superheating and supercooling in gases~\cite{Callen1985},
dynamic bistability, and relaxation into phase coexistence.

\begin{figure}[t]
\includegraphics[width=\columnwidth]{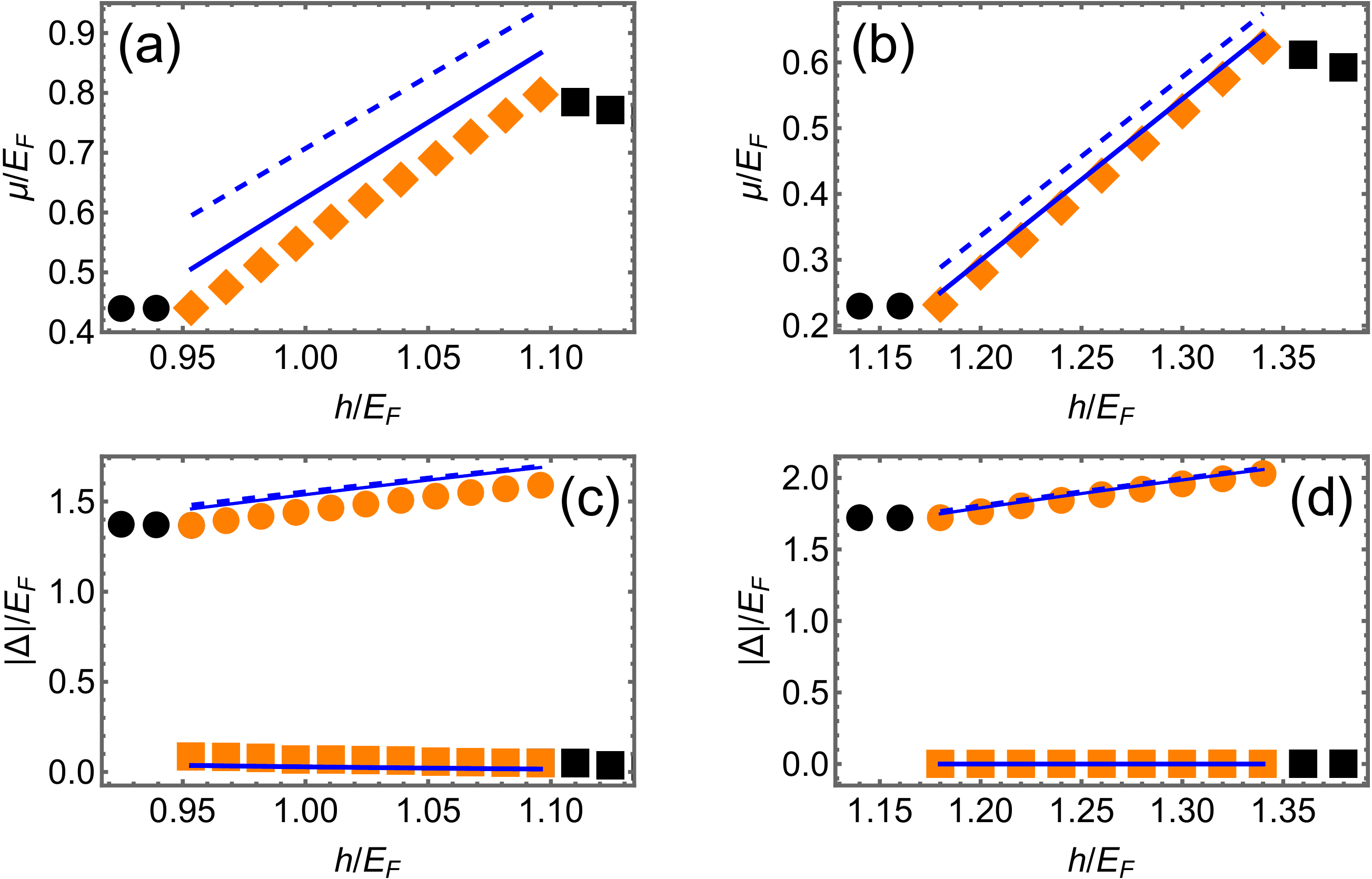}
\caption{\label{fig:phaseCoexApp}%
Comparison of the approximate analytical expressions
[Eqs.~(\ref{eqs:muDelAna}a--c), shown as solid blue curves] for
the chemical potential $\mu$ and pair-potential magnitudes
$|\Delta_\mathrm{s}|$ and $|\Delta_\mathrm{w}|$ in the
phase-coexistence region with exact numerical results (shown by
symbols with same conventions as in Fig.~\ref{fig:Coex}).
Panels (a) and (c) are for the same situation as depicted in
panels (b) and (e) of Fig.~\ref{fig:Coex} ($\lambda = 0.35\,
E_\mathrm{F}/k_\mathrm{F}$, $E_\mathrm{b} = 1.0 \,
E_\mathrm{F}$). The results shown in panels (b) and (d) have
been calculated assuming $\lambda = 0.20\, E_\mathrm{F}/
k_\mathrm{F}$ and $E_\mathrm{b} = 1.5 \, E_\mathrm{F}$. For
illustration, dashed blue curves show results for the $\lambda
= 0$ case according to exact analytical formulae from
Ref.~\cite{He2008}.}
\end{figure}

The $h$-dependence of the chemical potential is observed to be
quite strong in the phase-coexistence regime, interpolating
approximately linearly between the typically very different
single-phase values $\mu(h_<)$ and $\mu(h_>)$. In contrast, the
magnitudes of the pair potentials $|\Delta_\mathrm{s}|$ and
$|\Delta_\mathrm{w}|$ vary much more slowly as $h$ is tuned
across the phase-coexistence region. Finiteness of
$|\Delta_\mathrm{w}|$ is a direct consequence of finite
spin-orbit coupling, and its magnitude is enhanced
monotonically as $\lambda$ increases. Such features and further
trends observed in the numerically obtained self-consistent
values for the chemical potential and pair-potential magnitudes
can be discussed more systematically and quantitatively based
on our approximate analytical results. Here we focus on the
situation where the first-order phase transition coincides with
the topological transition, i.e., when the weak superfluid is a
TSF. The middle and right columns of panels in
Fig.~\ref{fig:Coex} pertain to examples for such a scenario.
For the case of weak spin-orbit coupling that is realized when
the condition $\lambda k_\mathrm{F}\ll \mathrm{min}
\{E_\mathrm{F}, h\}$ holds for the relevant range of Zeeman
energies within the phase-coexistence region, it is possible to
derive approximate analytical expressions that generalize those
obtained previously~\cite{He2008} when $\lambda=0$. To leading
order in explicit small-$\lambda$ corrections, we find (see the
Appendix for details of the derivation)
\begin{subequations}\label{eqs:muDelAna}
\begin{eqnarray}\label{eq:approxMu}
\mu &\approx& \frac{\sqrt{2}h - E_\mathrm{b}}{2 - \sqrt{2}} -
\frac{\lambda^2 k_\mathrm{F}^2}{2\sqrt{2} E_\mathrm{F}} \left(
\frac{\sqrt{2} + 1}{\sqrt{2} - 1} \frac{E_\mathrm{b}}{2h} - 1
\right) \,\, , \nonumber\\ \\[5pt] \label{eq:approxDs}
|\Delta_\mathrm{s}| &\approx& \sqrt{E_\mathrm{b}^{(\lambda)}
\left( E_\mathrm{b}^{(\lambda)} + 2\mu \right)} \quad , \\[5pt]
\label{eq:approxDw}
|\Delta_\mathrm{w}| &\approx& \ee^\frac{5}{2}\, \frac{\ee^{- \left[
\frac{2 E_\mathrm{F}}{h + \mu} \ln \left( \frac{2
h}{E_\mathrm{b}} \right) \right] \frac{h^2}{\lambda^2
k_\mathrm{F}^2}}}{\lambda k_\mathrm{F}/h} \,
\frac{\sqrt{E_\mathrm{F}(\mu + h)}}{\ee^{\frac{h - \mu}{h +
\mu} \ln\left( \frac{2h}{h - \mu}\right)}} \,\,\,\, .
\end{eqnarray}
\end{subequations}
For compactness, we introduced the effective
spin-orbit-coupling-renormalized binding energy
\begin{equation}\label{eq:lambdaEb}
E_\mathrm{b}^{(\lambda)} = E_\mathrm{b} + \frac{\lambda^2
k_\mathrm{F}^2}{2 E_\mathrm{F}}\frac{E_\mathrm{b}}{E_\mathrm{b}
+ \mu\big|_{\lambda=0}}
\end{equation}
that is relevant for the strong-superfluid phase. The form
(\ref{eq:lambdaEb}) of $E_\mathrm{b}^{(\lambda)}$ captures the
general tendency of spin-orbit coupling to strengthen
pairing~\cite{Chen2012,He2012a}, but the influence of that on
the magnitude of $|\Delta_\mathrm{s}|$ is counteracted by a
concomitant decrease in $\mu$ as per Eq.~(\ref{eq:approxMu}).
Figure~\ref{fig:phaseCoexApp} illustrates the applicability of
the approximate analytical expressions from
(\ref{eqs:muDelAna}a--c) by comparing them with the exact
numerical values for $\mu$, $|\Delta_\mathrm{s}|$ and
$|\Delta_\mathrm{w}|$.

Our results for $\mu$ [Eq.~(\ref{eq:approxMu})] and the
pair-potential magnitude $|\Delta_\mathrm{s}|$ of the
strong-superfluid phase [Eq.~(\ref{eq:approxDs})] exhibit only
small corrections due to finite spin-orbit coupling. This is
expected because these quantities should smoothly recover known
results~\cite{He2008} for the $\lambda=0$ situation. In
particular, the dependence of $|\Delta_\mathrm{s}|$ on
$\lambda$ is extremely weak (as was previously also observed in
the $h=0$ limit~\cite{Chen2012,He2012a}), and it becomes a
function of $h$ only implicitly through its dependence on
$\mu$. Thus the expression (\ref{eq:approxDs}) together with
(\ref{eq:approxMu}) captures the trend of $|\Delta_\mathrm{s}|$
to increase approximately linearly with $h$ while varying only
insignificantly with $\lambda$ as long as $\lambda \lesssim
\sqrt{E_\mathrm{F} h}/k_\mathrm{F}$.
 
As the existence of the weak-superfluid phase is predicated on
spin-orbit coupling being finite, $|\Delta_\mathrm{w}|$ depends
materially on $\lambda$. This is embodied by the analytical
expression given in Eq.~(\ref{eq:approxDw}), which was derived
for the situation when the weak superfluid is topological;
i.e., for $h > h_\mathrm{c} > \mu$, assuming also spin-orbit
coupling to be small. According to this formula, which
constitutes one of our main results, $|\Delta_\mathrm{w}|$
vanishes in a singular fashion for $\lambda\to 0$, as surmised
in a previous numerical analysis~\cite{Zhou2011}. The
comparatively weak $h$ dependence of $|\Delta_\mathrm{w}|$ is
also reproduced by the intricate functional form of
Eq.~(\ref{eq:approxDw}).

\begin{figure}[t]
\includegraphics[width=\columnwidth]{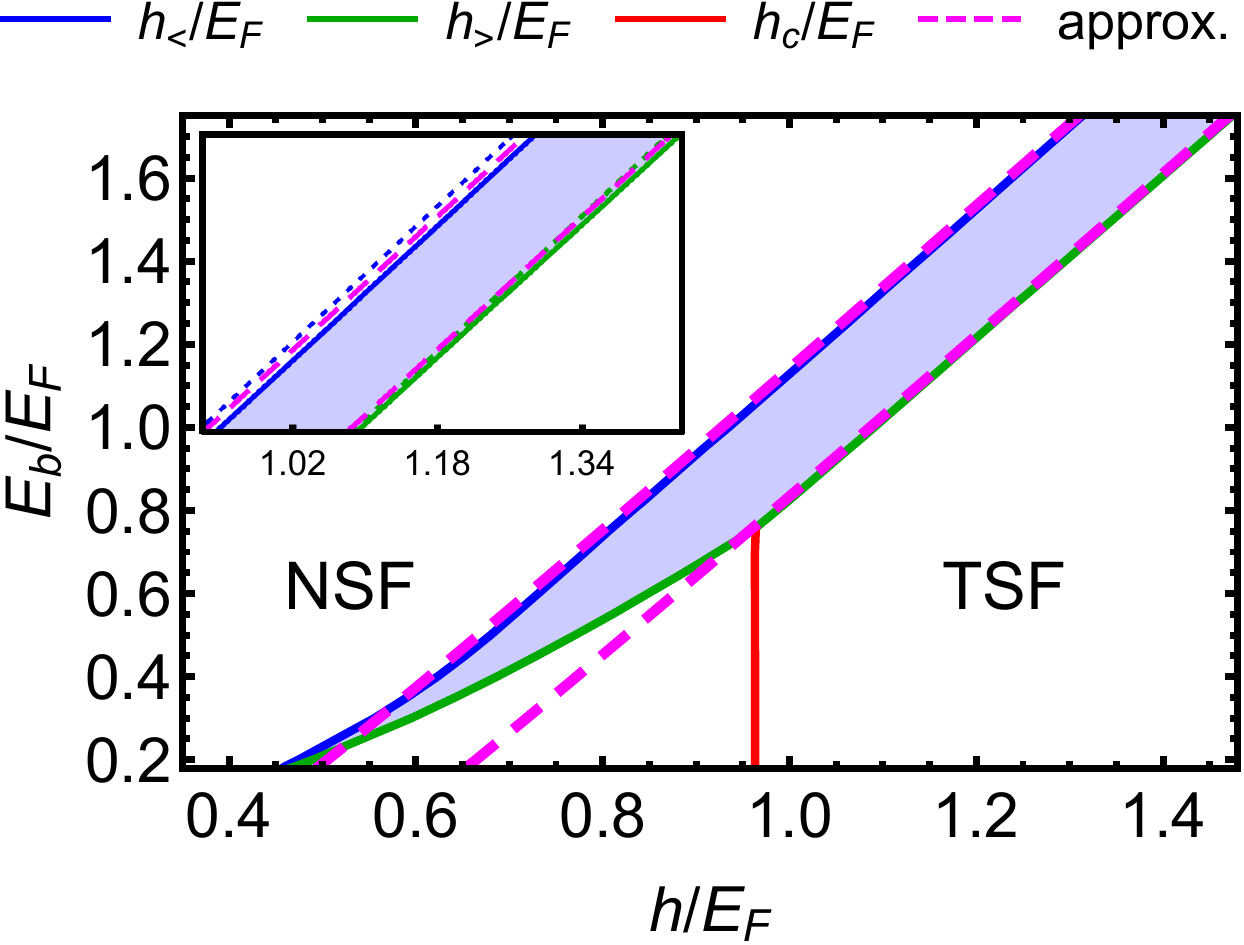}
\caption{\label{fig:hCritComp}%
Comparison of the approximate analytical expressions
Eqs.~(\ref{eqs:approxHcri}a,b) for the critical Zeeman energies
$h_<$ and $h_>$ (shown as long-dashed magenta curves) with
exact numerical results (shown as the solid blue and green
curves), for the case $\lambda = 0.25\, E_\mathrm{F}/
k_\mathrm{F}$. The red curve indicates $h_\mathrm{c}$ defined
in Eq.~(\ref{eq:topCrit}). Phase coexistence is between a TSF
and an NSF when $h_> > h_\mathrm{c}$. This is the regime for
which the analytical formulae were derived and where the
agreement is excellent. The inset shows results for a reduced
parameter range together with the $\lambda=0$ critical fields
plotted as the short-dashed blue and green curves.}
\end{figure}

The formula (\ref{eq:approxDw}) relies on the condition $2h >
E_\mathrm{b}$ being satisfied in the TSF--NSF phase-coexistence
region. This is indeed the case because $2 h_< > E_\mathrm{b}$
holds, as can be seen from both the exact numerical results and
the approximate analytical expressions for critical fields,
\begin{subequations}\label{eqs:approxHcri}
\begin{eqnarray}
h_< &\approx& (\sqrt{2} - 1)\,E_\mathrm{F} +
\frac{E_\mathrm{b}}{2} + \frac{\lambda^2 k_\mathrm{F}^2}{2E_\mathrm{F}}\, \frac{E_\mathrm{b} - (2-\sqrt{2})
E_\mathrm{F}}{E_\mathrm{b} + 2(\sqrt{2}-1)E_\mathrm{F}} \,\, ,
\nonumber \\ \\ h_> &\approx& (2 - \sqrt{2}) \, E_\mathrm{F} +
\frac{E_\mathrm{b}}{2} + \frac{\lambda^2 k_\mathrm{F}^2}{2
E_\mathrm{F}}\, \frac{E_\mathrm{b} -  \sqrt{2} E_\mathrm{F} }{E_\mathrm{b} + 2(2
- \sqrt{2}) E_\mathrm{F}} \,\, . \nonumber \\
\end{eqnarray}
\end{subequations}
The formulae (\ref{eqs:approxHcri}) are generalizations of the
previously obtained~\cite{He2008} $\lambda=0$ expressions for
the critical fields. Details of our derivation are given in the
Appendix, and Fig.~\ref{fig:hCritComp} illustrates the
applicability of Eqs.~(\ref{eqs:approxHcri}) by a comparison
with our exact numerical results. Although the leading
corrections to $h_<$ and $h_>$ due to finite spin-orbit
coupling are of the same order, their actual magnitudes turn
out to be quite different within the shown parameter range ---
relevant for $h_<$ but insignificant for $h_>$.

Our theoretical approach is based on BdG mean-field theory
whose applicability to low-dimensional systems is known to be
limited~\cite{Bertaina2011,Turlapov2017} to zero temperature
and weak-enough interactions. Our choice of parameters is
designed to reflect these limitations, ensuring in particular
the condition $\mu > 0$ for the BCS regime.

\section{\label{sec:concl}Conclusions}

We have performed a zero-temperature mean-field study of the
first-order superfluid-superfluid transition in a 2D spin-$1/2$
Fermi gas with attractive two-particle interactions in the
\textit{s}-wave channel and also subject to Zeeman spin
splitting and spin-orbit coupling. Both numerical and
analytical results are presented in Sec.~\ref{sec:PhaseCoex} to
characterize the regime where two superfluid phases with
different magnitudes of the \textit{s}-wave pair potential
coexist. The total fermion density $n$ is assumed to be fixed,
and we absorb density dependences of physical quantities by
using $k_\mathrm{F}\equiv\sqrt{2\pi n}$ and $E_\mathrm{F}\equiv
\hbar^2 k_\mathrm{F}^2/(2m)$ as wave-vector and energy scales,
respectively. Thus, relevant parameters controlling the system
properties are the dimensionless spin-orbit-coupling strength
$\lambda k_\mathrm{F}/E_\mathrm{F}$, Zeeman energy $h/
E_\mathrm{F}$, and the interaction strength parameterized in
terms of $E_\mathrm{b}/E_\mathrm{F}$, where $E_\mathrm{b}$ is
the two-particle binding energy in vacuum in the absence of
spin-orbit coupling and Zeeman spin splitting.
Figure~\ref{fig:Coex} shows results for situations where the
first-order transition is driven by changing $h/E_\mathrm{F}$,
keeping all other parameters fixed. We focus on this scenario
because increasing $h/E_\mathrm{F}$ above a critical value
$h_\mathrm{c}$ [see Eq.~(\ref{eq:topCrit})] is associated with
the establishment of a topological-superfluid (TSF) phase in
our system of interest. Above a critical value for
$E_\mathrm{b}/E_\mathrm{F}$, the topological transition gets
intertwined with the first-order phase transition (see
Fig.~\ref{fig:Diag}), and the TSF coexists with a
nontopological superfluid (NSF) over an extended range $h_< < h
< h_>$ of Zeeman energies. The boundary between spatial domains
occupied by different phases will generally be shaped by the
requirement of energy minimization and/or the form of external
potentials.

Coexistence of TSF and NSF phases arises in our system of
interest as a consequence  of the first-order phase transition
driven by raising the Zeeman energy $h$. It is directly linked
to the coexistence between superfluid and nonsuperfluid phases
in a polarized Fermi gas without spin-orbit coupling. In a
different context, TSF--NSF coexistence was proposed to occur
in harmonically trapped Fermi gases~\cite{Zhou2011,Yang2012a,
Liu2012,Liu2012a,Xu2014} where both the chemical potential and
the pair potential have an $r$-dependent profile, causing a
spatially varying critical Zeeman energy $h_\mathrm{c}(r)
\equiv \sqrt{\mu(r)^2 + |\Delta(r)|^2}$ for the second-order
topological transition. When the Zeeman energy $h$ is adjusted
to satisfy $h_\mathrm{c}(0) > h > h_\mathrm{c}(r_\mathrm{F})$,
where $r_\mathrm{F}$ is the trap's Thomas-Fermi radius, the
system consists of a central region where $h<h_\mathrm{c}(r)$,
which is thus an NSF, surrounded by an outer region where $h>
h_\mathrm{c}(r)$, thus realizing a TSF. Numerical BdG mean
field studies of 2D~\cite{Liu2012} and 1D~\cite{Liu2012a,
Xu2014} trap geometries have elucidated general features of the
regime where both TSF and NSF phases are present. There is
overall qualitative agreement between our results and those
presented in Refs.~\cite{Liu2012,Liu2012a,Xu2014}, even though
the physical origin of TSF--NSF phase coexistence is ostensibly
different in both. However, looking closely at the numerically
obtained pair-potential profiles (Fig.~2 in
Ref.~\cite{Liu2012a} and Fig.~S1 in the Supplemental Material
for Ref.~\cite{Xu2014}), one recognizes a quite sharp drop at
the inner edge (the TSF--NSF interface), which is flanked by
regions of basically constant pair-potential magnitudes. Such
behavior is more evocative of the first-order phase coexistence
considered in our work than a system divided into TSF and NSF
regions by virtue of a trap-induced continuous variation of
both chemical and pair potentials. This observation leads us to
surmise that the numerical studies from Refs.~\cite{Liu2012,
Liu2012a,Xu2014} encountered the first-order phase-coexistence
scenario in a finite-size system.

One of the intriguing characteristics of TSFs is the presence
of unconventional quasiparticle excitations at boundaries. The
TSF--NSF interface in the coexistence regime discussed in the
present work constitutes such a boundary. Our results can feed
directly into a theoretical description of the interface
modeled as an SS$^\prime$ junction where the pair potential
changes abruptly between its configuration in the two
coexisting homogeneous phases~\cite{Thompson202x}. This
approach can provide useful guidance for the experimental
investigation and potential manipulation of Andreev bound
states at the TSF--NSF interface. Pursuing such avenues for
in-depth study within well-controlled ultracold-atom setups can
be expected to yield crucial insights about the, at this point,
equally intriguing and elusive,
Majorana quasiparticles~\cite{vanHeck2017,Prada2020,
Dmytruk2020,Pan2021}.

\begin{acknowledgments}
This work was supported by the Marsden Fund Council (contract
no.\ VUW1713), from New Zealand government funding managed by
the Royal Society Te Ap\=arangi. The authors thank Philip
Brydon for useful comments.
\end{acknowledgments}

\appendix*

\section{\label{app:Anal}Derivation of approximate analytical
expressions for chemical potential and pair-potential
magnitudes in the TSF--NSF-coexistence regime}

Our analytical approach is based on approximating the true
mean-field ground-state energy
$\mathcal{E}_\mathrm{gs}^\mathrm{(MF)}(|\Delta|, \mu)$ in the
phase-coexistence region by expressions
$\mathcal{E}_\mathrm{gs}^{(\mathrm{s})}(|\Delta|, \mu)$ and
$\mathcal{E}_\mathrm{gs}^{(\mathrm{w})}(|\Delta|, \mu)$ that
accurately capture the functional dependence on the
pair-potential magnitude around its degenerate minima
$|\Delta_\mathrm{s}|$ and $|\Delta_\mathrm{w}|$, respectively.
We also consider only the weak-spin-orbit-coupling limit, which
means practically that nonleading-order corrections in
$\lambda k_\mathrm{F}/h$ and $\lambda k_\mathrm{F}/
E_\mathrm{F}$ are neglected.

We start by determining $\mathcal{E}_\mathrm{gs}^{(\mathrm{s})}
(|\Delta|, \mu)$, which describes the strong-superfluid part of
the system, i.e., the one with the larger pair-potential
magnitude $|\Delta_\mathrm{s}|$. As phase coexistence requires
a sufficiently large $E_\mathrm{b}\ge
E_\mathrm{b}^{(\mathrm{c})}$~\cite{Thompson2020}, it turns out
that the condition $h < |\Delta_\mathrm{s}|$ holds in
situations where the first-order transition also constitutes
the topological transition. Using this condition alongside the
one for small $\lambda$ in calculating the expression
(\ref{eq:gsEn}) for the mean-field ground-state energy, we
obtain to leading order in both small quantities ($\lambda
k_\mathrm{F}/E_\mathrm{F}$ as well as $h/|\Delta|$)
\begin{widetext}
\begin{equation}\label{eq:GSstrongApp}
\mathcal{E}_\mathrm{gs}^{(\mathrm{s})}(|\Delta|,\mu) = n\,
\frac{|\Delta|^2}{4 E_\mathrm{F}} \left[ \ln \frac{\sqrt{\mu^2
+ |\Delta|^2} - \mu}{E_\mathrm{b}} - \frac{\mu}{\sqrt{\mu^2 +
|\Delta|^2} - \mu} - \frac{1}{2} - \frac{\lambda^2
k_\mathrm{F}^2}{E_\mathrm{F}}\, \frac{1 + \frac{h^2}{3
|\Delta|^2} \, \frac{\sqrt{\mu^2 + |\Delta|^2} +
\mu}{\sqrt{\mu^2 + |\Delta|^2}}}{\sqrt{\mu^2 + |\Delta|^2} -
\mu} \right] \,\, .
\end{equation}
In the limit $\lambda \to 0$, Eq.~(\ref{eq:GSstrongApp})
reduces to the familiar $h$-independent ground-state energy
for the 2D Fermi gas with finite Zeeman spin splitting when it
is in its unpolarized superfluid phase~\cite{Randeria1990,
He2008}. For $h\to 0$, (\ref{eq:GSstrongApp}) also recovers the
previously obtained~\cite{Chen2012} leading-order correction to
the ground-state energy due to spin-orbit coupling. For the
purposes of our following calculations, the term $\propto h$ in
$\mathcal{E}_\mathrm{gs}^{(\mathrm{s})}(|\Delta|,\mu)$ will be
neglected, as it constitutes a further small correction to a
contribution that is already small in the parameter $\lambda
k_\mathrm{F}/E_\mathrm{F}$.

We now proceed to obtain
$\mathcal{E}_\mathrm{gs}^{(\mathrm{w})}(|\Delta|, \mu)$, which
has its minimum at the pair-potential magnitude
$|\Delta_\mathrm{w}|$ of the weak-superfluid part of the
phase-coexistence state. In the absence of spin-orbit coupling,
this phase would be a normal (nonsuperfluid) Fermi
gas~\cite{He2008}. This implies that $|\Delta_\mathrm{w}|$
vanishes for $\lambda\to 0$ and, thus, will generally be small
in the weak-spin-orbit-coupling limit. Focusing on situations
where the weak superfluid phase is topological, we assume $h >
\sqrt{\mu^2 + |\Delta|^2}$ together with $\lambda k_\mathrm{F}
\ll h$ when evaluating the r.h.s.\ of Eq.~(\ref{eq:gsEn}),
which yields the result
\begin{eqnarray}\label{eq:GSweakApp}
\mathcal{E}_\mathrm{gs}^{(\mathrm{w})}(|\Delta|, \mu) =&&
\frac{n}{4 E_\mathrm{F}} \left\{ |\Delta|^2 \left[ \ln \frac{h
+ \sqrt{h^2 - |\Delta|^2}}{E_\mathrm{b}} - \frac{1}{2} \right]
- h\, \sqrt{h^2 - |\Delta|^2} - \mu^2 - 2 h\mu \right.
\nonumber \\[5pt] && \left. - \, \frac{\lambda^2
k_\mathrm{F}^2}{h E_\mathrm{F}} \left[ \frac{(\mu + h)^2}{2} +
\mu\left( \sqrt{h^2 - |\Delta|^2} - h \right) + 2 |\Delta|^2
\left( \frac{3}{4} + \frac{\mu}{\sqrt{h^2 - |\Delta|^2}}
\right) \right] \right. \nonumber\\[5pt] && \left. + \,
\frac{\lambda^2 k_\mathrm{F}^2}{h E_\mathrm{F}}\,
\frac{|\Delta|^2}{2} \left[\left( 1 + \frac{\mu}{\sqrt{h^2 -
|\Delta|^2}} \right) \ln \frac{\lambda k_\mathrm{F}
|\Delta|}{E_\mathrm{F}^2} + \left( 1 - \frac{\mu}{\sqrt{h^2 -
|\Delta|^2}} \right)\ln \frac{2\, \sqrt{h^2 -
|\Delta|^2}}{\sqrt{h^2 - |\Delta|^2} - \mu} \right. \right.
\nonumber \\[5pt] && \hspace{4.5cm} \left. \left. +\,  \left(
1 + \frac{\mu}{\sqrt{h^2 - |\Delta|^2}} \right) \ln
\frac{E_\mathrm{F}^{3/2}}{\sqrt{(h^2 - |\Delta|^2) (\mu +
\sqrt{h^2 - |\Delta|^2})}} \right] \right\} \,\, . \quad
\end{eqnarray}
\end{widetext}
The result (\ref{eq:GSweakApp}) generalises the formula for the
ground-state energy of a fully polarised ($h > \mu$) 2D Fermi
gas to the situation of having a finite $|\Delta|$ due to
$\lambda\ne 0$.

The expressions for $|\Delta_\mathrm{s,w}|$ are determined from
the minimum-ground-state-energy conditions
\begin{equation}\label{eq:MinGSEnergyApp}
\left. \frac{\partial
\mathcal{E}_\mathrm{gs}^{(\mathrm{s,w})}}{\partial |\Delta|}
\right|_{|\Delta|\to |\Delta_\mathrm{s,w}|} = 0 \quad ,
\end{equation}
yielding Eqs.~(\ref{eq:approxDs}) and (\ref{eq:approxDw}) when
only leading-order contributions in small quantities $\lambda
k_\mathrm{F}/E_\mathrm{F}$, $\lambda k_\mathrm{F}/h$ and
$|\Delta_\mathrm{w}|/h$ are retained. The chemical potential
$\mu$ is derived from the phase-coexistence condition
(\ref{eq:coexCond}), written within our analytical approach as
\begin{equation}\label{eq:CoexCondApp}
\mathcal{E}_\mathrm{gs}^{(\mathrm{s})}(|\Delta_\mathrm{s}|,\mu)
= \mathcal{E}_\mathrm{gs}^{(\mathrm{w})}(|\Delta_\mathrm{w}|,
\mu) \quad .
\end{equation}
With leading-order accuracy in small $\lambda$ and
approximating $\mathcal{E}_\mathrm{gs}^{(\mathrm{w})}
(|\Delta_\mathrm{w}|,\mu)\approx \mathcal{E}_\mathrm{gs}^{(
\mathrm{w})}(0,\mu)$, (\ref{eq:CoexCondApp}) can be expressed
as
\begin{equation}\label{eq:CoexCondExplApp}
\frac{1}{2} \left( 2\mu + E_\mathrm{b} + \frac{\lambda^2
k_\mathrm{F}^2}{E_\mathrm{F}} \right)^2 = (\mu + h)^2 \left(
1 + \frac{\lambda^2 k_\mathrm{F}^2}{2 h E_\mathrm{F}} \right)
\,\, .
\end{equation}
Solving Eq.~(\ref{eq:CoexCondExplApp}) for $\mu$ gives
Eq.~(\ref{eq:approxMu}). Figure~\ref{fig:phaseCoexApp}
illustrates the validity of the analytical expressions obtained
here for the chemical potential and pair-potential magnitudes
in the phase-coexistence regime.

The expressions $\mathcal{E}_\mathrm{gs}^{(\mathrm{s,w})}$ can
also be utilized to find the chemical potentials
$\mu_\mathrm{s}$ and $\mu_\mathrm{w}$ in the homogeneous
nontopological (strong) and topological (weak) superfluid
phases that adjoin the phase-separation region. These can then
be used to find analytical expressions for the critical Zeeman
energies $h_<$ and $h_>$, by equating $\mu$ as given in
Eq.~(\ref{eq:approxMu}) with $\mu_\mathrm{s}$ and
$\mu_\mathrm{w}$, respectively. For implementing this
procedure~\cite{He2008}, we apply the fixed-density conditions
\begin{equation}\label{eq:fixedNumb}
\left. \frac{\partial \mathcal{E}_\mathrm{gs}^{(\mathrm{s,w})}}
{\partial \mu}\right|_{\mu\to \mu_\mathrm{s,w}} = -n \quad .
\end{equation}
Using first Eq.~(\ref{eq:GSstrongApp}) for
$\mathcal{E}_\mathrm{gs}^{(\mathrm{s})}$ in
(\ref{eq:fixedNumb}), inserting Eq.~(\ref{eq:approxDs}) for
the pair-potential magnitude, and collecting leading-order
terms in the spin-orbit-coupling strength yields the
$h$-independent result~\cite{He2013,Brand2018}
\begin{equation}\label{eq:muS}
\mu_\mathrm{s} \approx E_\mathrm{F} - \frac{E_\mathrm{b}}{2} -
\frac{\lambda^2 k_\mathrm{F}^2}{2 E_\mathrm{F}}
\end{equation}
that coincides with the expression derived previously for the
spin-orbit-coupled 2D superfluid at $h=0$~\cite{Chen2012}.
Turning to determining $\mu_\mathrm{w}$, we use
Eq.~(\ref{eq:GSweakApp}) for
$\mathcal{E}_\mathrm{gs}^{(\mathrm{w})}$ in
(\ref{eq:fixedNumb}) and insert Eq.~(\ref{eq:approxDw}) for
$|\Delta|$. Keeping only leading-order $\lambda$-dependent
corrections (which implies neglecting any terms with
exponentials that vanish faster than any power law as
$\lambda\to 0$), we find
\begin{equation}\label{eq:muW}
\mu_\mathrm{w} \approx 2E_\mathrm{F} - h - \frac{\lambda^2
k_\mathrm{F}^2}{h} \quad ,
\end{equation}
consistent with previous results given for the chemical
potential of the topological 2D Fermi
superfluid~\cite{Brand2018}. Equating $\mu_\mathrm{s}$ from
Eq.~(\ref{eq:muS}) and $\mu_\mathrm{w}$ from
Eq.~(\ref{eq:muW}), respectively, with $\mu$ from
Eq.~(\ref{eq:approxMu}) yields Eqs.~(\ref{eqs:approxHcri}) for
the critical Zeeman energies. Figure~\ref{fig:hCritComp}
illustrates the validity of the analytical description within
its region of applicability, i.e., when the weak superfluid
phase is a TSF.

%

\end{document}